# Striations in electronegative capacitively coupled radio-frequency plasmas: effects of the pressure, voltage, and electrode gap

Yong-Xin Liu[1], Ihor Korolov[2], Edmund Schüngel[3], You-Nian Wang[1], Zoltán Donkó[2], and Julian Schulze[4,5]

1) Key Laboratory of Materials Modification by Laser, Ion, and Electron Beams (Ministry of Education), School of Physics, Dalian University of Technology, Dalian 116024, China
2) Institute for Solid State Physics and Optics, Wigner Research Centre for Physics, Hungarian Academy of Sciences, H-1121 Budapest, Konkoly-Thege Miklós str. 29-33, Hungary
3) Evatec AG, Hauptstrasse 1a, CH-9477 Truebbach, Switzerland
4) Institute for Electrical Engineering, Ruhr-University Bochum, 44801 Bochum, Germany
5) Department of Physics, West Virginia University, Morgantown, West Virginia 26506-6315, USA

Capacitively coupled radio-frequency (CCRF) $CF_4$ plasmas have been found to exhibit a self-organized striated structure at operating conditions, where the plasma is strongly electronegative and the ion-ion plasma in the bulk region (largely composed of $CF_3^+$ and $F^-$ ions) resonates with the excitation frequency. In this work we explore the effects of the gas pressure, the RF voltage, and the electrode gap on this striated structure by Phase Resolved Optical Emission Spectroscopy and Particle-In-Cell/Monte Carlo Collisions simulations. The measured electronic excitation patterns at different external parameters show a good general agreement with the spatio-temporal plots of the ionization rate obtained from the simulations. For a fixed driving frequency the minima of the $CF_3^+$ and $F^-$ ion densities (between the density peaks in the bulk) are comparable and independent of other external parameters. However, the ion density maxima generally increase as a function of the pressure or RF voltage, leading to the enhanced spatial modulation of plasma parameters. The striation gap (defined as the distance between two ion density peaks) is approximately inversely proportional to the pressure, while it exhibits a weak dependence on the RF voltage and the electrode gap. A transition between the striated and non-striated modes can be observed by changing either the pressure or the RF voltage; for 13.56 MHz and 18 MHz driving frequencies we present a phase diagram as a function of the pressure and voltage amplitude parameters.

# I. INTRODUCTION

Capacitively coupled radio frequency (CCRF) discharges operated in electronegative gases, e.g., $O_2$, $Cl_2$, $CF_4$, $SF_6$, etc. have been widely employed for plasma etching and deposition in semiconductor manufacturing and for various other types of material surface processing applications [1-3]. A more in-depth understanding of the fundamental processes in CCRF plasmas can improve the performance of plasmas sources used in these practical applications, which could ultimately create enormous social benefits. During the past few years, many researchers studied electronegative CCRF plasmas [4-16]. Investigations of the charged species' dynamics are most important to obtain a detailed understanding of the fundamental mechanisms of plasma generation as a basis for improving the performance of such plasma sources. For example, the electron power absorption mode of the plasma determines the electron energy distribution function that has a direct effect on the gas-phase chemical reactions, e.g., ionization, excitation, dissociative attachment processes, etc. These microscopic processes ultimately determine the concentrations and spatial distributions of charged species and radicals [17-22]. As to the ion dynamics, the acceleration and collisionality of various ionic species within the sheath region defines their energy distribution functions at the substrates, which play a key role in the interaction of the plasma with material surfaces [23-27].

In low-pressure electropositive CCRF discharges, e.g., in Ar gas, electrons gain energy primarily at the edges of the oscillating sheaths ($\alpha$-mode) [19, 20, 28, 29]. At relatively high pressure and/or high applied voltage, the plasma is primarily sustained by the ionization inside the sheath region caused by secondary electrons emitted from the electrodes due to ion bombardment ($\gamma$-mode) [30]. In these two modes, the electron power deposition, as well as the ionization/excitation rate is typically high around the sheath edges and low in the plasma bulk [4]. By contrast, in plasmas operated in electronegative gases, such as $CF_4$, $O_2$, $SiH_4$, etc., electrons can also gain energy inside the bulk region [5-11]. This is made possible by the drift and ambipolar electric fields inside the bulk plasma and at the edges of the collapsing sheaths, which effectively heat electrons [drift-ambipolar (DA) mode] [12-14]. The high drift and ambipolar electric fields are primarily caused by the presence of negative ions in the bulk, whose densities can be comparable to those of positive ions [15]. In most previous studies, the (positive and negative) ions in this drift field have generally been considered to be at rest (being non-responsive to the drift electric field, which oscillates at the driving frequency, e.g., 13.56 MHz, due to their heavy masses), in contrast to the mobile electrons that instantaneously follow the rapidly alternating electric field.

However, when the ion plasma frequencies, $\omega_\pm = \sqrt{e^2 n_\pm / \varepsilon_0 m_\pm}$ , where $e$ is the elementary charge, $n_\pm$ is the (positive or negative) ion density, $\varepsilon_0$ is the permittivity in vacuum, and $m_\pm$ are the ion masses, becomes comparable to, or higher than the driving frequency, positive and negative ions may respond to the RF alternating electric field, leading to the formation of space charges [31, 32]. The electric field caused by the space charges enhances or attenuates the local drift electric field in the bulk, resulting in at first a slight spatial modulation of the electric field and of the ion densities. The spatially modulated electric field then reinforces the response of the positive and negative ions to the RF electric field by pushing them towards the locations of the ion density maxima. Consequently, the space charges, as well as the striated electric field and the ion density maxima are enhanced due to this positive feedback. The effect is self-amplified until a periodic steady state of spatially modulated electric field profiles is established. The striated electric field results in a spatial modulation of the electron power absorption and, consequently, of the electron-impact excitation/ionization rate. Self-organized striated structures of the plasma emission–established this way–have been observed in CCRF $CF_4$ plasmas by Phase Resolved Optical Emission Spectroscopy (PROES) and their formation was analyzed and understood by Particle in Cell / Monte Carlo Collision (PIC/MCC) simulations [31]. It is quite obvious that the physical origin of the striations in electronegative CCRF discharges is different from those observed previously in electropositive plasmas [33-46]. For example, one well-known scenario is the striated structure of DC glow discharges, wherein ion-acoustic or ionization waves are considered to be the origin of these features [33-36]. Striations can also occur in inductively coupled plasmas [37-39], atmospheric pressure CCRF $He/H_2O$ discharges [40], atmospheric pressure plasma jets [41-43], plasma display panels [44], and plasma clouds in the ionosphere [45, 46], etc. In all of these systems, the appearance of striations is associated with the electron kinetics.

The operation modes of RF plasmas are very sensitive to the external control parameters, and mode transitions induced by changes of these control parameters have been observed and extensively studied experimentally and via numerical methods [4, 28, 47-52]. Godyak *et al*. [47] studied experimentally the transition between the *α* and *γ* modes in argon and helium plasmas, and observed a sharp rise in the electron density and a drop in the electron temperature by increasing the driving current, because of the efficient generation of low-energy electrons via ionization caused by secondary electrons in the "high-voltage" mode. By using PIC/MCC simulation Denpoh *et al.* [48, 49] and Proshina *et al.* [50] observed a transition from the bulk heating mode (DA mode) into the sheath heating mode (α-mode) with decreasing pressure and/or increasing RF voltage in $CF_4$ capacitive discharges. With the same particle-based kinetic approach, Yan *et al.* [51, 52] compared the spatio-temporal distributions of the electron heating, as well as the ionization dynamics in electropositive argon and in electronegative silane capacitive discharges. They observed heating mode transitions induced by changing the pressure, the power, and the driving frequency. The mode transitions induced by these parameters were later confirmed by a comparative study comprising PROES measurements and PIC/MCC simulations. [4, 28] The underlying physics of the mode transitions in electronegative CCRF discharges is primarily correlated with the variation of the electronegativity $\xi$ as a function of the external parameters [12, 13, 28, 53]. Here $\xi = \bar{n}_-/\bar{n}_e$, where $\bar{n}_-$ and $\bar{n}_e$ are the spatial averages of the negative ion and electron density, respectively, inside the plasma bulk.

Similarly, the external control parameters are also expected to have a significant influence on the striated structures in electronegative CCRF discharges. We have already reported the effect of the driving frequency on striations in CCRF $CF_4$ discharges over a frequency domain of 4 – 27.12 MHz, at fixed values of the other parameters (i.e., the pressure, voltage amplitude, and electrode gap) [32]. The striations were found to occur within a wide frequency range. In the striated (STR) mode, the ion density minima were found to exhibit an approximately quadratic increase with the driving frequency, while the maximum ion density was found to be abnormally enhanced due to the ions being focused into the striations by the spatially modulated electric field inside the bulk. By increasing the driving frequency, the discharge was found to switch from the STR mode into the "regular" DA mode. However, it is still unclear how the other external parameters, such as the neutral gas pressure, the RF voltage, and the electrode gap, affect the properties of the striations, e.g., the spatially modulated profiles of the electric field, ionization/excitation rate, ion densities, etc., in electronegative CCRF plasmas.

In this work, following the description of the experimental setup and the simulation method in section 2, in section 3 we first present a systematic study of the effects of the gas pressure (*p*), voltage amplitude (*ϕ*), and electrode gap (*L*) on the striations at a fixed driving frequency of 8 MHz, based on PROES measurements and PIC/MCC simulations. At this frequency changes of these parameters do not induce mode transitions. Subsequently, we analyze the mode transition between the STR and DA modes induced by changes of the working pressure and the voltage amplitude at driving frequencies of 13.56

MHz and 18 MHz via PIC/MCC simulations. Finally, conclusions are given in section 4.

## II. EXPERIMENT AND SIMULATION

### A. Experimental setup

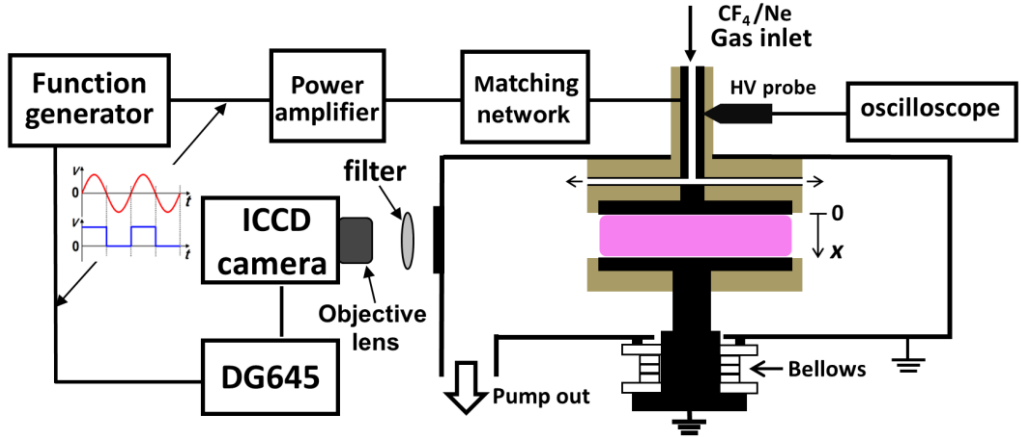

FIG. 1. Schematic of the CCRF plasma chamber, supplemented with a phase resolved optical emission spectroscopy diagnostic system.

The experimental setup of the CCP reactor and the diagnostic system used for the PROES measurements are shown in figure 1 (also described in Ref. 32). The plasma is produced in $CF_4$ with a 5% admixture of Ne as a probe gas for PROES, between two parallel circular stainless-steel electrodes with 10 cm diameter. The gas mixture, with a total gas flow rate of 25 sccm (standard cubic centimeter per minute), is first introduced through a central tube along the axis of the upper electrode and then spreads outwards through 8 outlets (45 degrees apart) into the chamber according to figure 1. The gas is pumped out of the reactor through the side port of the chamber by a turbomolecular pump. The bottom electrode and the chamber wall are grounded. The electrode gap can be adjusted by changing the height of the lower electrode. A sinusoidal RF signal of a two-channel function generator (Tektronix AFG 3252C) is connected to a power amplifier (AR, Model 1000A225), whose output is applied to the top electrode via a matching network. Due to the geometric asymmetry of the reactor, the discharge is asymmetric as well, resulting in the generation of a DC self-bias voltage, whose absolute value changes as a function of the discharge conditions. The voltage waveform is measured close to the powered electrode by a high-voltage probe (Tektronix P6015A) and is acquired with a digitizing oscilloscope (LeCroy Waverunner). All measured voltage waveforms are sinusoidal.

The other channel of the function generator is connected to a pulse delay generator (SRS INC., Model DG645) that triggers synchronously an intensified charge-coupled device (ICCD) camera (Andor iStar DH734) for the PROES measurements. The ICCD camera is equipped with an objective lens and an interference filter to detect the spatio-temporal emission intensity of a specifically chosen optical transition (at 585 nm) of the neon tracer gas. From the light emission measurements, that are performed in a sequence through the RF period, the spatio-temporal electron-impact excitation rate from the ground state into the $N_e 2p_1$-state is calculated based on a collisional-radiative model (for more details see Refs. 54, 55).

Our experiments are carried out at different pressures (40 Pa – 120 Pa), voltage amplitudes (100 V – 500 V), and electrode gaps (1.5 cm – 2.5 cm), while the driving frequency is fixed either at 8 MHz (sections 3.1 – 3.3) or at 13.56 MHz/18 MHz (section 3.4). The former value of 8 MHz is chosen in sections 3.1 – 3.3 to avoid mode transitions as a function of $p$, $\phi$, and $L$, but to study the effects of these parameters on the STR mode. 13.56 MHz and 18 MHz are chosen in section 3.4 to allow mode transitions to occur as a function of $p$ and $\phi$.

## B. Simulation model and method

Our numerical studies are based on a bounded plasma Particle in Cell simulation code, complemented with a Monte Carlo type treatment of collision processes (PIC/MCC), see, e.g., Ref. 56. The code considers one spatial (axial) coordinate and is three-dimensional in the velocity space. The electrodes of the discharge are assumed to be plane and parallel. The cross sections of electron-$CF_4$ collision processes are taken from Ref. 57, with the exception of the electron attachment processes (producing $CF_3^-$ and $F^-$ ions), for which we use data from Ref. 58. We use an extensive set of electron-impact collision processes, however, disregard many of the products created in these reactions. The only products considered are $CF_3^+$, $CF_3^-$, and $F^-$ ions, which play the most important role in $CF_4$ discharge plasmas. These ions can participate in various ion-molecule reactive processes, as well as in elastic collisions [59-61]. The ion-ion recombination rate coefficients are taken from ref 50, while the rate for the electron-$CF_3^+$ recombination process is from ref 49.

In the simulations we assume a gas temperature of 350 K. We neglect ion-induced emission of secondary electrons from the electrodes and assume that electrons are reflected at the electrodes [62] with a probability of 0.2. More details of our model, tables and graphical representations of the cross sections can be found in Ref. 63. In the simulations we use 600 grid points to resolve properly the inter-electrode gap and between 15 000 and 30 000 time steps within an RF period, depending on the driving frequency, to resolve properly the temporal dynamics of all plasma species and to fulfill the relevant stability criteria of the numerical method.

The simulations are carried out for parameter sets matching the experimental conditions. It should be kept in mind, however, that due to the asymmetry of the experimental electrode configuration, the measured voltage waveforms exhibit a large negative self-bias for some conditions, which cannot be accounted for by the simulations. In all cases the amplitudes of the measured voltage waveforms are used as input for the PIC/MCC simulations. We also note that comparisons will be made between the spatio-temporal distributions of the excitation rate measured in the experiments and the ionization rate calculated in the simulations. This approach is justified by the fact that both of these processes represent the dynamic behavior of electron groups with similar energies, due to their comparable threshold energies (the excitation threshold of the $2p_1$ state of neon is ~19 eV, while the ionization threshold of $CF_4$ molecules is ~16 eV).

## III. RESULTS

In sections 3.1 – 3.3 the effects of the neutral gas pressure $p$, the voltage amplitude $\phi$, and the electrode gap $L$ on the striations are discussed at a fixed driving frequency of 8 MHz. At this driving frequency the plasma remains in the STR mode for all choices of $p$, $\phi$, and $L$ studied here. In contrast to that, at fixed driving frequencies of 13.56 MHz and 18 MHz mode transitions are induced by tuning these parameters. This scenario is studied in section 3.4.

Before discussing the effects of the parameter variations, however, we briefly recall the main physical mechanisms responsible for the emergence of striations. The positive ion–negative ion bulk plasma formed under highly electronegative conditions may be susceptible to a resonance with the driving RF signal at $\omega_+^2 + \omega_-^2 \geq \omega_{RF}^2$ (where $\omega_{+,-}$ is the positive/negative ion plasma frequency and $\omega_{RF}^2$ is the driving frequency), whenever the ion densities reach a critical value, and the positive and negative ions may respond to the RF alternating electric field. [31,32] The major ion species in a $CF_4$ plasma are $CF_3^+$ and $F^-$. Thus, we disregard the minority species $CF_3^-$ in the frame of this resonance condition. The motion of the ions generates space charges, where density gradients are present, which enhance/attenuate the local electric field. The ions created in collisions of electrons accelerated in the high bulk electric field with the neutral molecules are focused by the local fields into the striations. This process generates enhanced local ion density peaks, where a balance of ionization and losses via recombination determines the maximum density values. In the periodic steady state of the discharge strong spatial modulations of all plasma characteristics (electric field, charged particle densities, electron impact excitation/ionization rates, etc.) are found [32]. In the following the peculiarities of these features as a function of the control parameters $p$, $\phi$, and $L$ are analyzed.

## A. Pressure variation

First we investigate the effects of changing the gas pressure on the striated structure of various plasma characteristics. Figure 2 shows the pressure dependence of the spatio-temporal distributions of the measured electron-impact excitation rate (first column), the calculated ionization rate (second column), the electric field (third column), the net charge density (fourth column), and the electron power absorption rate (fifth column) at a fixed driving frequency of $f = 8$ MHz, voltage amplitude of $\phi = 300$ V, and electrode gap of $L = 1.5$ cm. Generally, there is a good agreement between the experimentally observed spatio-temporal distributions of the excitation and the ionization rate obtained from the simulation. As $p$ increases, the gap between the excitation/ionization rate maxima (defined as "striation gap") clearly decreases both in the experiment and in the simulation. Similarly, a decreasing distance between the maxima of the spatially modulated electric field, net space density, and electron power absorption rate with increasing pressure can be observed in the respective panels of figure 2. At $p = 40$ Pa the experimental excitation patterns show contributions of energetic electrons created by the fast expansion of the sheath at the powered electrode, and a combined α–D/A mode of operation, while the simulation results indicate operation in the STR mode. This discrepancy (spatial asymmetry) is caused by the high negative self-bias generated in the experiment due to the geometric asymmetry of the plasma reactor. At p = 40 Pa, the DC self bias $\phi_b$ is about -145 V. Its absolute value is found to decrease as a function of $p$ and, meanwhile, the measured excitation patterns within the two halves of the RF cycle become more similar, indicating the establishment of a more symmetrical discharge in the experiment. Besides, as $p$ increases, the intensive excitation and ionization regions both in the experiment and in the simulation expand towards the plasma bulk and the spatial modulation of the excitation/ionization rate, the electric field, and the electron power absorption rate are gradually enhanced. The agreement between the measured spatio-temporal excitation rate and the simulated ionization rate is improved with increasing gas pressure.

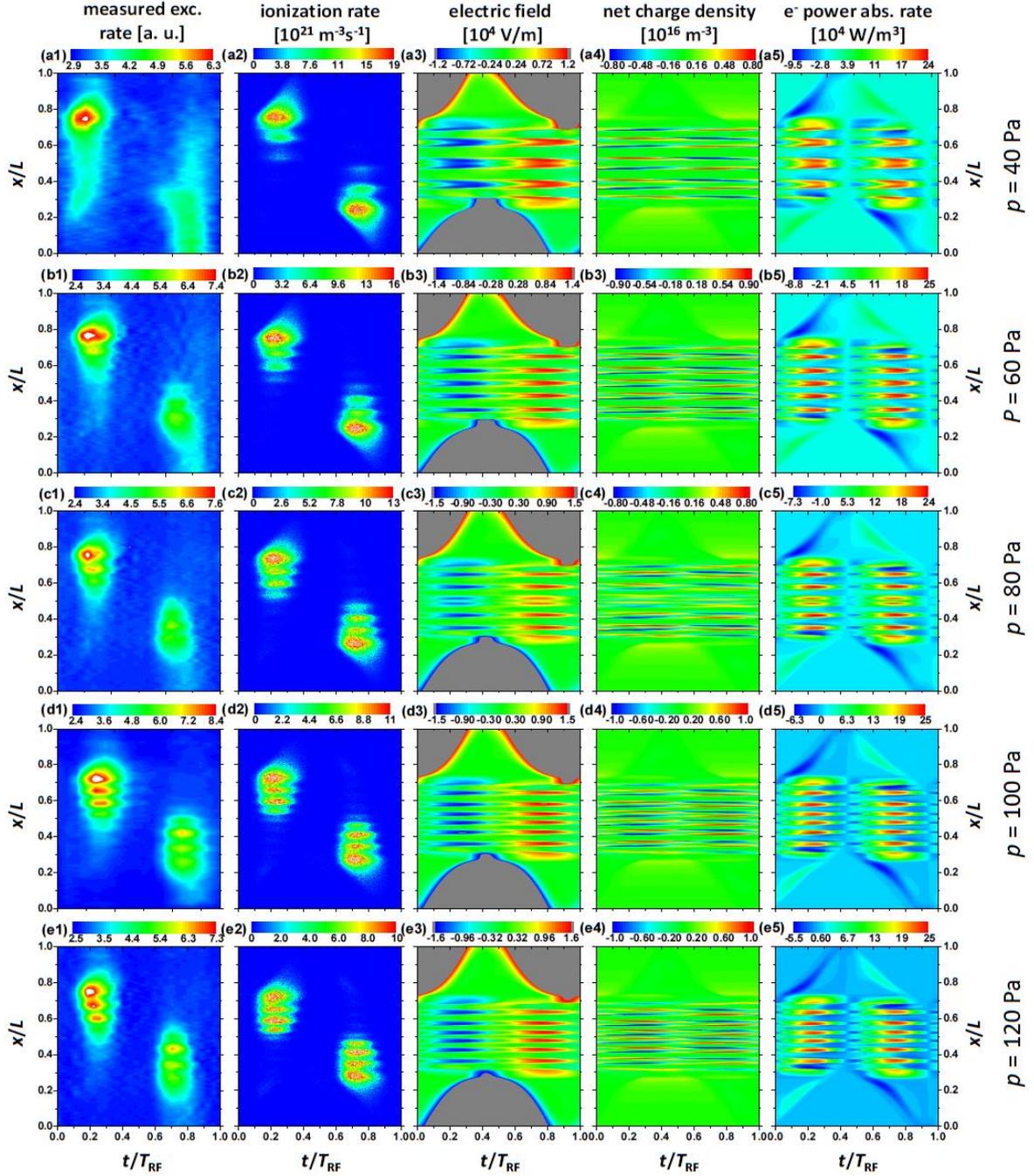

FIG. 2. PROES results: electron-impact excitation rate at different pressures (first column). PIC /MCC simulation results: the spatio-temporal plots of the electron-impact ionization rate (second column), electric field (third column), net charge density (fourth column) and electron power absorption rate (fifth column) at different pressures. Discharge conditions: driving frequency $f$ = 8 MHz, voltage amplitude $\phi$ = 300 V, electrode gap $L$ =1.5cm. The powered electrode is situated at $x$ = 0, while the grounded electrode is located at $x/L$ = 1.

Figures 3(a-e) show simulation results for the axial profiles of the time-averaged charged species densities in the pressure range of 40 – 120 Pa. It can be seen that the $CF_3^+$ and $F^-$ ion densities exhibit comb-like profiles and that the striation gap (i.e., the distance between two ion density maxima) decreases as $p$ increases. For all values of $p$, the minima of the densities of the major ionic species ($CF_3^+$ and $F^-$) are comparable in the bulk, as the minimum densities of the different ions are determined by the requirement to fulfill the resonance condition (see above). [32]

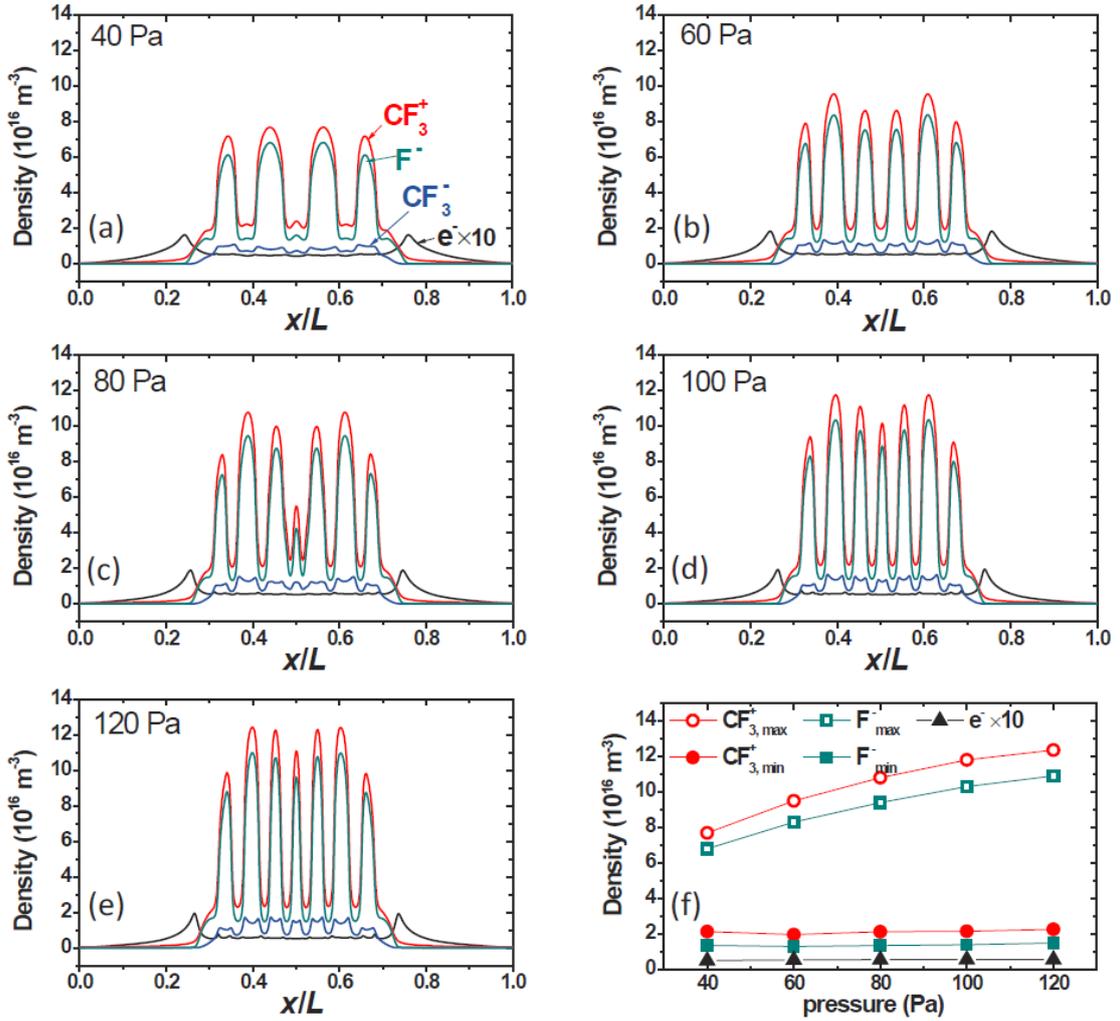

FIG. 3. (a-e) Axial profiles of the time-averaged $CF_3^+, F^-, CF_3^-$ ion and electron densities at pressure $p$ = 40, 60, 80, 100, and 100 Pa. (f) Maximum and minimum of the $CF_3^+$ (circles) and $F^-$ (rectangles) ion densities and the space-averaged electron density $\bar{n}_e$ (triangles) inside the bulk region a function of $p$. (PIC/MCC simulation results.) The electron density is multiplied by a factor of 10. The conditions are the same as in figure 2.

The maxima and minima of the $CF_3^+$ and $F^-$ ion densities and the space-averaged electron density ($\bar{n}_e$) over the bulk region as a function of $p$ are presented in figure 3 (f). It can be seen that the minima of the $CF_3^+$ and $F^-$ ion densities, as well as $\bar{n}_e$ are almost independent of $p$, while the maximum ion densities increase by a factor of ~1.6 with the increase of $p$. Consequently, at the positions of the density maxima the local electronegativity increases with $p$. It is worth noting that at the lowest pressure (40 Pa), the maximum ion density is much higher than the critical ion density (at which the resonance condition is met), which is close to the minimum of the ion density in figure 3 and, thus, within the pressure range covered here the striations are always present. [32]

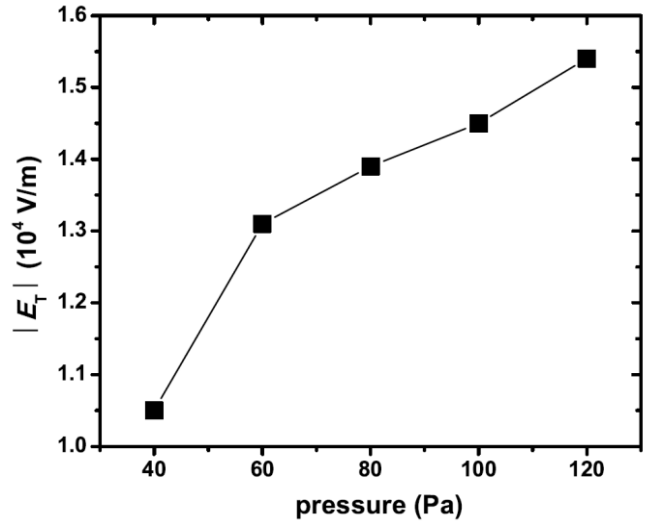

FIG. 4. Amplitude of the electric field oscillation $|E_T|$ at the location of the central ion density minimum as a function of p (PIC/MCC results). The conditions are the same as in figure 2.

Figure 4 illustrates the amplitude of the transient electric field oscillation $|E_T|$ as a function of $p$, close to the center of the bulk. It can be seen from figure 4 that $|E_T|$ increases monotonously with $p$. This is caused by a decrease of the electrical conductivity due to an increasing electron-neutral collision frequency: in order to drive a given current through the plasma a higher electric field is required at higher pressures. As analyzed in our previous work [32], such a transient electric field can focus the positive and negative ions into the locations of the local ion density peaks, leading to the enhancement of the maximum ion density and the reduction of the minimum ion density. So, at a higher pressure the stronger transient electric field can cause more significant spatial modulations of the ion density profiles and, consequently, of the ionization/excitation rates and other plasma parameters.

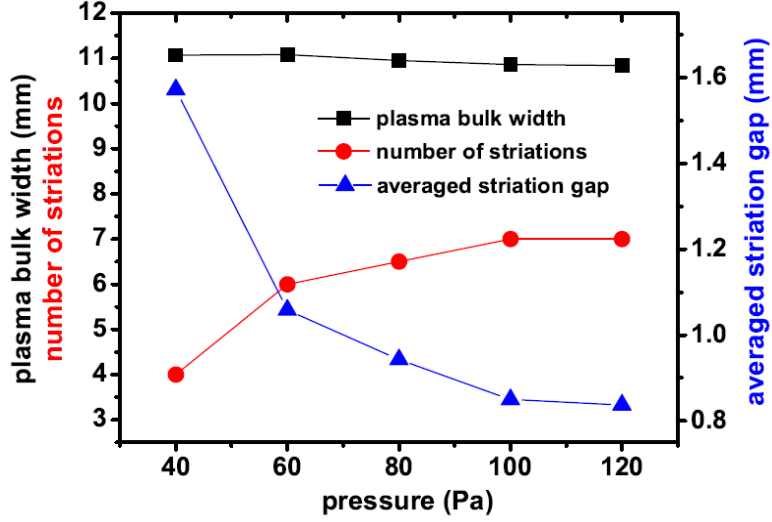

FIG.5. Simulation results for the plasma bulk width (black rectangles), number of ion density peaks (circles), and the average striation gap (triangles) as a function of *p*. The conditions are the same as in figure 2.

Figure 5 shows the plasma bulk width *b* as a function of *p*. The plasma bulk width *b* is determined as $b = L - s_{max}$, where *L* is the electrode gap and $s_{max}$ is the maximum width of the sheath. The sheath width is determined by Brinkmann's criterion. [64] The width of the bulk region exhibits a slight decrease with *p*, while the number of striations, also shown in figure 5, first increases rapidly, and then stabilizes at 7 at $p \geq 100$ Pa. So, the striation gap is approximately inversely proportional to *p,* as also displayed in figure 5.

The pressure-dependence of the electric field strength (see figure 4) in combination with the decrease of the striation width (see figure 3) as a function of *p* allows understanding the effect of the pressure on the striation gap by revoking the charged particle balance mechanisms. The positive and negative ions are created by electrons that are accelerated by the electric field between the striations. These ions are focused onto the closest striation, where they accumulate and recombine with each other. This way the balance between particle generation and losses is expected to define the length of the striation gap. Electrons cross multiple striations during one time interval of high electric field within one RF period (see figure 2). Meanwhile, they gain energy, when they move through regions of high electric field between two ion density maxima, and they loose energy via collisions, when they move through regions of depleted electric field at the positions of the striations [32]. The width of each striation and, thus, the distance over which electrons loose energy decreases as a function of pressure, since the amplitude of the ion displacement, $A^* = \frac{eE_0/\mu}{\omega_{RF}\nu}$ , decreases as a function of $p$ similarly to the decrease of the electron mean free path [32] . Here, $E_0$ is the electric field, $\mu$ is the reduced mass of the positive and negative ions, and $\nu$ is the ion collision frequency. Thus, the collisional electron energy loss, that occurs, when an electron crosses a striation, remains approximately constant as a function of pressure. In combination with the increase of $|E_T|$ in between the striations electrons can gain sufficient energy to balance the recombination losses by ionization in the vicinity of each striation over shorter striation gaps.

**B. Voltage variation**

In figure 6 we present the voltage amplitude dependence of the spatio-temporal distributions of the measured electron-impact excitation and the calculated ionization rate, as well as of the electric field, the net charge density and the electron power absorption rate for $f$ = 8MHz, $p$ = 100 Pa, $L$ = 1.5 cm. The comparison between the measured spatio-temporal excitation and the simulated ionization rate shows good agreement. At the low voltage amplitude of $\phi$ = 100 V the spatial modulation of the light emission is relatively weak. More pronounced patterns emerge in the ionization rate (figure 6(a2)). A clear striated structure in the experiment develops at $\phi \geq$ 200 V. With an increasing $\phi$ the striated structure becomes gradually more pronounced and the region with intense excitation/ionization concentrates towards the edge of the collapsing sheath. This is caused by a decrease of the bulk width as a function of $\phi$ due to larger sheaths (see figure 9). During one half of an RF period electrons cross multiple striations and are accelerated by the striated electric field. Their energy increases as they move from the expanding to the collapsing sheath edge. Increasing $\phi$ and, thus, decreasing the bulk width shrinks the region, where electrons are accelerated. Therefore, at a given position in the bulk their energy is decreased as a function of $\phi$, and they reach high-enough energies to cause ionization only at the edge of the bulk at the collapsing sheath edge for higher values of $\phi$. At the highest voltage amplitude values ($\phi$ = 400 V and 500 V) traces of the $\alpha$ power deposition mode are seen in the experimental data for the excitation rate (figure 6(d1, e1)) as faint features near the edge of the expanding sheath at the powered e1ectrode. These features are not observed at the grounded side. The different behaviors at the two sides of the discharge can be attributed to the geometric asymmetry of the plasma source. The DC self bias acquires its highest magnitude of $\phi_b \approx$ -150 V at ϕ = 500 V, and its absolute value is found to decrease with decreasing $\phi$. The self bias is almost zero at ϕ = 100 V, indicating a more symmetric discharge.

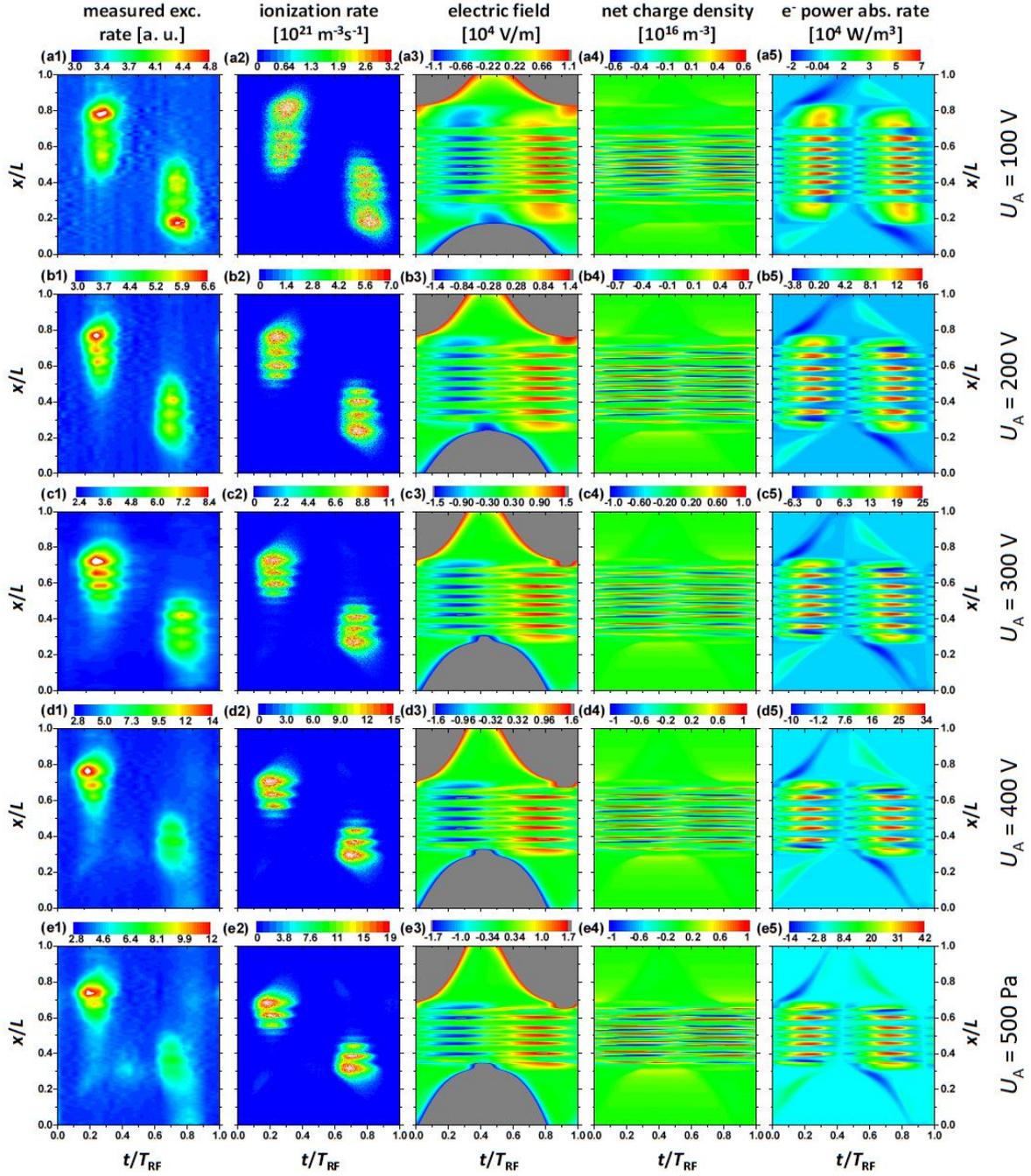

FIG. 6. PROES results: electron-impact excitation rate at different voltage amplitudes (first column). PIC/MCC simulation results: the spatio-temporal plots of the electron-impact ionization rate (second column), electric field (third column), net charge density (fourth column) and electron power absorption rate (fifth column) at different pressures. Discharge conditions: driving frequency $f$ = 8 MHz, $p$ = 100 Pa, electrode gap $L$ =1.5cm. The powered electrode is situated at $x = 0$, while the grounded electrode is located at $x/L = 1$.

Figures 7(a-e) show the axial profiles of the time-averaged charged species densities in the voltage amplitude range of 100-500 V. For all values of $\phi$, the $CF_3^+$ and $F^-$ ion densities exhibit comb-like profiles. Similar to the results in figure 3, the minima of the $CF_3^+$ and $F^-$ ion densities are not only comparable in the bulk, but are also almost independent of $\phi$, as it can be clearly seen from figure 7(f), which presents the maxima and minima of the $CF_3^+$, $F^-$ ion densities and $n_e$ as a function of $\phi$. By contrast, as $\phi$ increases from 100 to 500 V, the maximum $CF_3^+$ and $F^-$ ion densities generally increase by a factor of $\approx 2.2$, while $n_e$ increases by a factor of $\approx 3.8$, suggesting a slight decrease of the electronegativity due to the enhanced sheath heating at higher voltages. We note that at the lowest voltage (100 V), the maximum ion density is much higher than the critical ion density and, thus, the discharge is always striated [32].

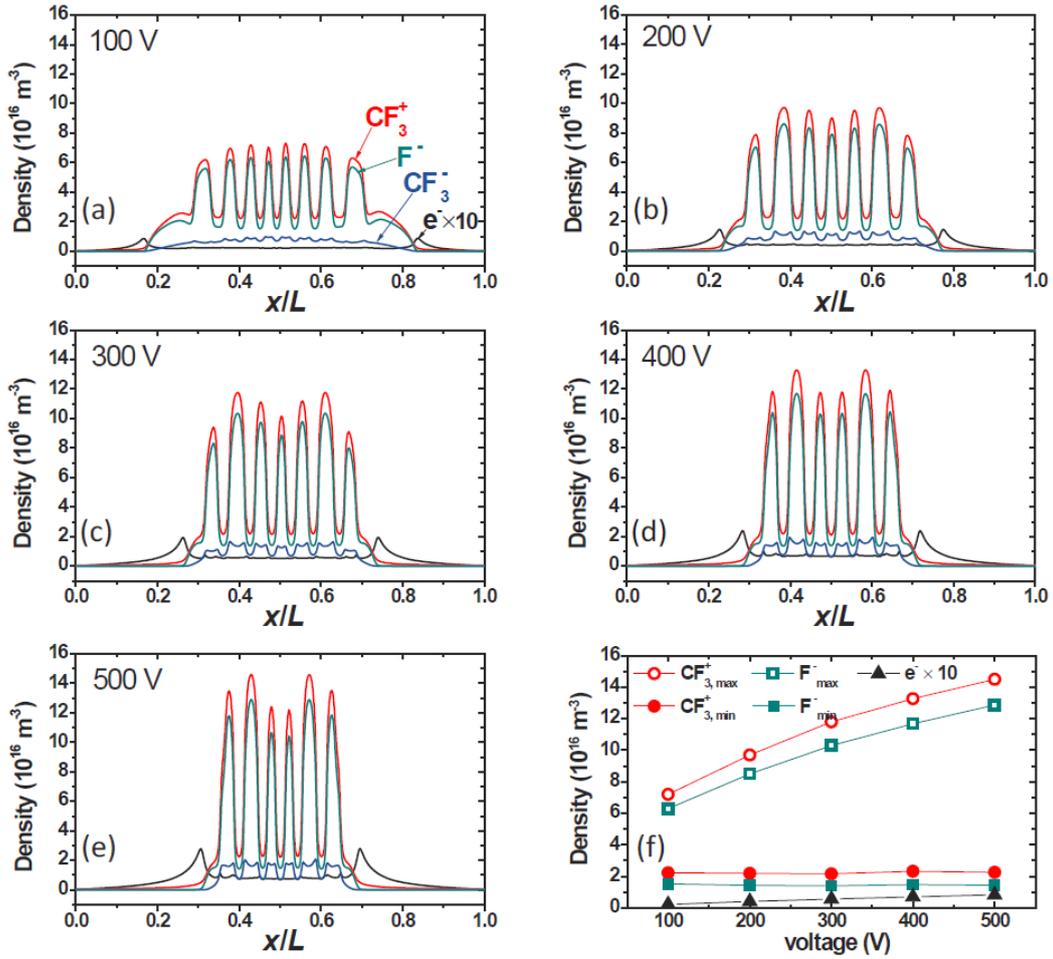

FIG. 7. PIC/MCC simulation results for (a-e) the axial profiles the time-averaged $CF_3^+$, $F^-$, $CF_3^-$ ion and electron densities at voltage amplitudes $\phi$ = 100, 200, 300, 400, and 500 V. (f) The maximum and minimum of the $CF_3^+$ (circles) and $F^-$ (rectangles) ion densities and $\bar{n}_e$ (triangles) inside the bulk

region as a function of $\phi$. The electron density is multiplied by a factor of 10. The conditions are the same as in figure 6.

In figure 8, we show the amplitude of the transient electric field oscillation $|E_T|$ at the location of the central ion density minimum, as a function of $\phi$. $|E_T|$ increases monotonically as a function of $\phi$, which explains the gradually enhanced spatial modulation of the plasma parameters with increasing $\phi$, shown in figure 6. This is caused by the fact that a higher current must be driven through the plasma at higher driving voltage amplitudes.

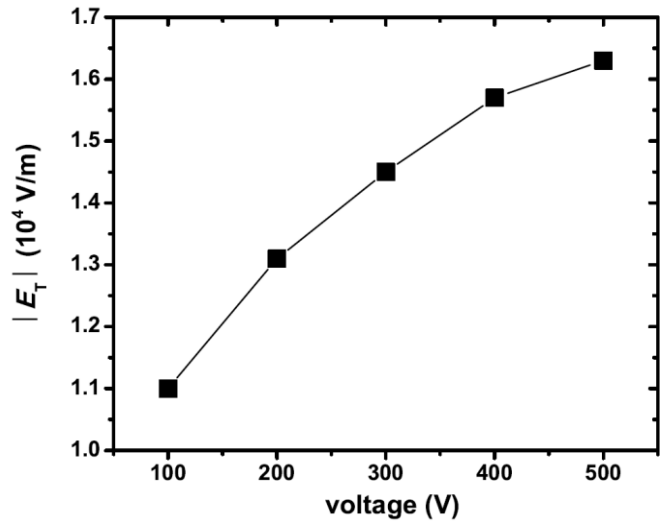

FIG. 8. PIC/MCC simulation results for the amplitude of the electric field oscillation $|E_T|$ at the location of the central ion density minimum as a function of $\phi$. The conditions are the same as in figure 6.

The dependence of the striation gap together with the number of striations, and the plasma bulk width on $\phi$ is displayed in figure 9. We see that the bulk plasma width, as well as the number of striations generally decrease as a function of $\phi$. So, we see an almost constant striation gap for different voltage amplitudes, as shown in figure 9. The striation width remains approximately constant as a function of $\phi$ similar to the ion displacement, A*. Thus, for a given pressure the collisional energy loss of a single electron, while it crosses a single ion density maximum, remains approximately constant as a function of the voltage. As $|E_T|$ increases as a function of $\phi$ and the striation gap remains constant, electrons can gain more energy due to their acceleration by the transient electric field between two striations at higher voltages and a balance of ionization and recombination can be maintained at higher ion densities such as observed in figure 7.

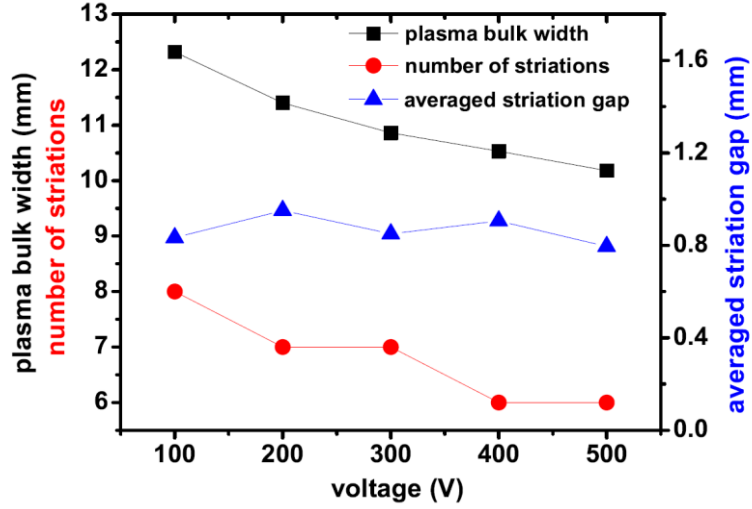

FIG. 9. PIC/MCC simulation results for the plasma bulk width (black rectangles), number of ion density peaks (circles), and the average striation gap (triangles) as a function of $\phi$. The conditions are the same as in figure 6.

## C. Gap variation

Figure 10 shows the spatio-temporal distributions of the measured electron-impact excitation and the calculated ionization rate, the electric field, the net charge density, and the electron power absorption rate at different electrode gaps under the conditions $f = 8$ MHz, $p = 100$ Pa, and $\phi = 300$ V. Note that the height of the vertical axis is proportional to the real gap length. By comparing the measured spatio-temporal excitation rate (first column) to the simulated ionization rate (second column), a good agreement is achieved. It can be seen that the striation gap does not change as a function of $L$. As $L$ increases, the spatial modulation of the measured excitation rate, the calculated ionization rate, and other plasma parameters is gradually weakened due to the reduction of the electric field inside the bulk (see figure 12).

At small electrode gaps, e.g., $L = 1.5$ cm, the intensive excitation/ionization concentrates near the edge of the collapsing sheath for each half of one RF period. As $L$ is enlarged, the intensive excitation/ionization region expands towards the edge of the opposite expanding sheath edge. When electrons are accelerated towards the collapsing sheath edge by the electric field in the bulk, it takes a certain distance for them to travel to reach the threshold energy for the excitation/ionization. Thus, for shorter gaps excitation/ionization occurs only at the collapsing sheath edge, while it is also present in the bulk for larger gaps. In the experiment we find that as $L$ is enlarged, the discharge gets more asymmetric (the absolute value of the DC self bias increases from 100 V to 130 V when $L$ increases from 1.5 cm to 2.5 cm) and this leads to the emergence of excitation patterns caused by the fast sheath expansion at the powered side (see figures 10(d1) and 10(e1)). A similar ionization pattern ($\alpha$-mode) can also be observed in the spatio-temporal plot of the ionization rate at large electrode gaps in the simulation (see figures 10(d2) and 10(e2)). In addition, we find that at $L = 1.5$ and $1.75$ cm, the local maximum of the net space charge is comparable over the entire bulk region, while the net space charge exhibits a maximum at the axial center at large gaps, i.e., $L \geq 2.0$ cm. This is due to the variation of the axial profile of the ion densities with $L$, as shown in figure 11(a–e).

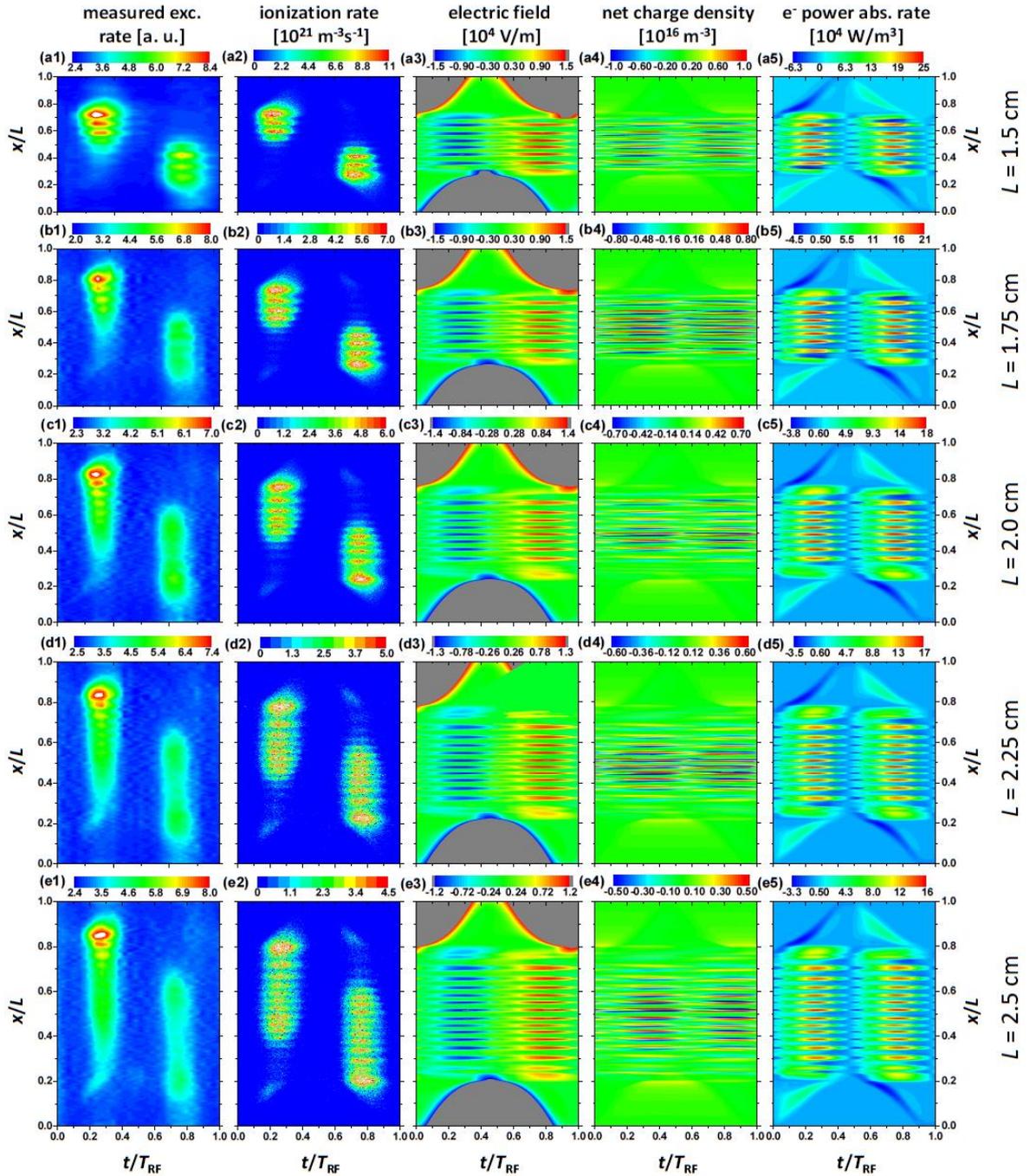

FIG. 10. (color online). PROES results: electron-impact excitation rate at different electrode gaps (first column). PIC/MCC simulation results: the spatio-temporal plots of the electron-impact ionization rate (second column), electric field (third column), net charge density (fourth column) and electron power absorption rate (fifth column) at different electrode gaps. Note that the height of the vertical axis is made in proportional to the real gap length. Discharge conditions: driving frequency $f$ = 8 MHz, $p$ = 100 Pa, and $\phi$ = 300 V. The powered electrode is situated at $x = 0$, while the grounded electrode is located at $x/L = 1$.

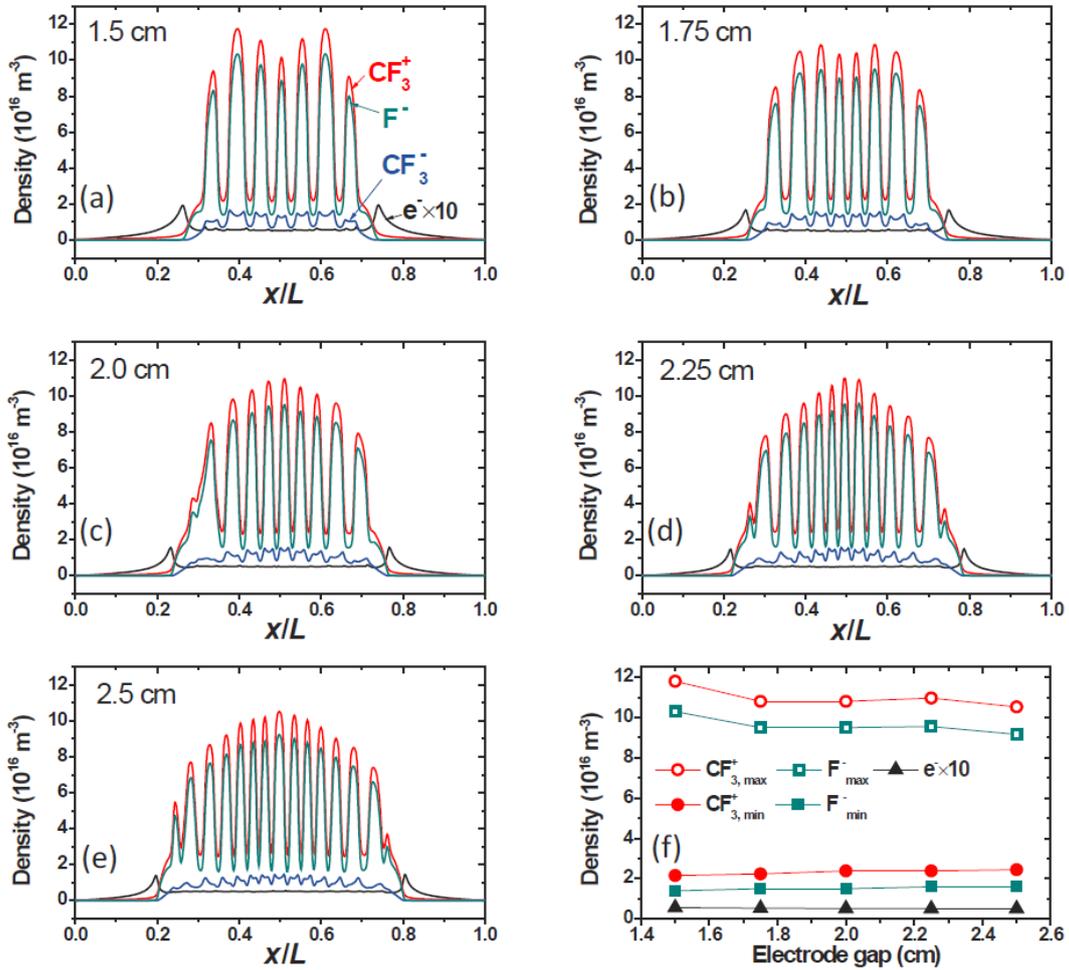

FIG. 11. (a-e) Axial profiles the time-averaged $CF_3^+, F^-, CF_3^-$ ion and electron densities at electrode gaps $L = 1.5, 1.75, 2.0, 2.25,$ and $2.5$ cm. (f) Maximum and minimum of the $CF_3^+$ (circles) and $F^-$ (rectangles) ion densities and $\bar{n}_e$ (triangles) inside the bulk region as a function of $L$. (PIC/MCC simulation results.) The electron densities is multiplied by a factor of 10. The conditions are the same as in figure 10.

Figure 11(a–e) shows the axial profiles of the time-averaged charged species densities at different electrode gaps, i.e., $L$ = 1.5–2.5 cm. The $CF_3^+$ and $F^-$ ion densities exhibit comb-like profiles. Similar to the results shown in figures 3 and 7, the minima of the $CF_3^+$ and $F^-$ ion densities remain constant as a function of $L$ in the bulk for the same reason given before. [32] However, for different electrode gaps, the maximum $CF_3^+$ and $F^-$ ion densities occur at different axial locations. To be specific, at $L$ = 1.5 and 1.75 cm, the ion densities show saddle-like axial profiles, with the maximum ion densities occurring at the positions $x/L \approx 0.4$ and $x/L \approx 0.6$, while the axial profiles are center-high at $L \geq 2.0$ cm (see figures 11(c–e)). This is because at small gaps the intensive ionization concentrates at the region close to the sheath edge, while it spreads all over the entire bulk region at large gaps (see the second column in figure 10). The maxima and minima of the $CF_3^+$ and $F^-$ ion densities, as well as $n_e$ as a function of $L$ are presented in figure 11(f). We observe that the minimum ion densities are independent of $L$, while with increasing $L$ the maximum ion densities as well as $n_e$ exhibit a slight drop, indicating almost unchanged electronegativity at different electrode gaps.

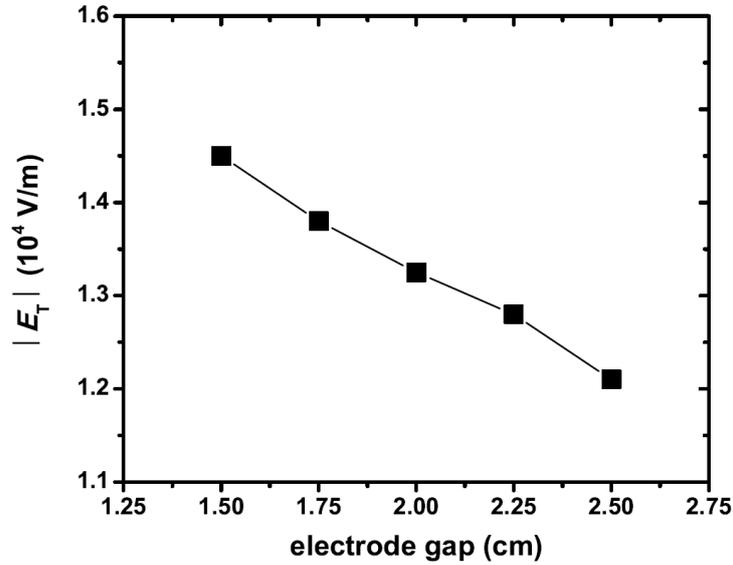

FIG. 12. Amplitude of the electric field oscillation $|E_T|$ at the location of the central ion density minimum as a function of $L$. (PIC/MCC simulation results.) The conditions are the same as in figure 10.

Figure 12 shows the amplitude of the transient electric field $|E_T|$ at the locations of the central ion density minimum as a function of $L$. $|E_T|$ decreases linear as a function of $L$, which leads to a gradually weakened spatial modulation of the plasma parameters (see figure 10). For a given $L$ and $\phi$ the potential drop across the plasma bulk is approximately constant, and consequently the electric field strength will generally decrease with the increase of *L* (or with the increase of the bulk width , see figure 13), resulting in the weakened spatial modulation of the electric field, as well as other plasma parameters.

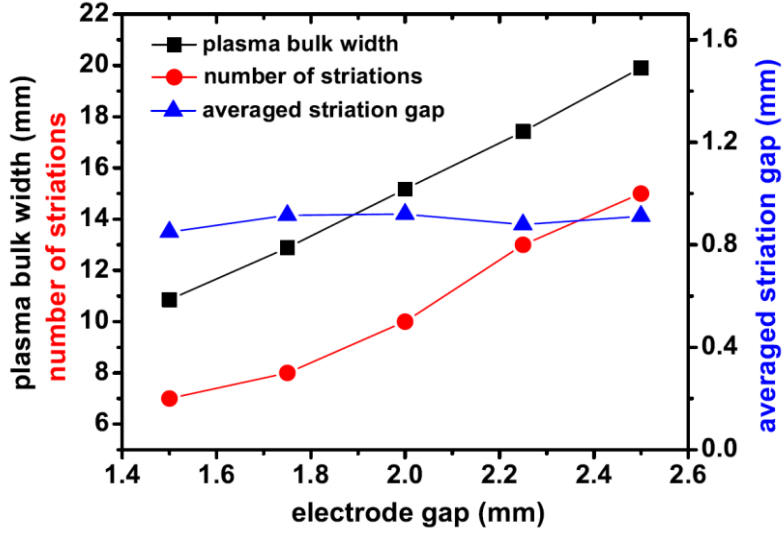

FIG. 13. PIC/MCC simulation results for the plasma bulk width (black rectangles), number of ion density peaks (circles), and the average striation gap (triangles) as a function of *L*. The conditions are the same as in figure 10.

The dependences of the striation gap, the number of striations, and the plasma bulk width on *L* are displayed in figure 13. We see that the bulk plasma width, as well as the number of striations increases linearly with *L*, leading to an almost constant striation gap at different electrode gaps (triangles in figure 13).

### D. Mode transition and phase diagram

In this section we discuss transitions between the DA and the STR modes induced by changing the pressure and the voltage amplitude. To map the behavior of the plasma over the ($\phi$, $p$) plane we have carried out simulations with voltage amplitudes between $\phi$ = 100 V and 400 V in 50 V steps, and at pressure values between $p$ = 30 Pa and 120 Pa in 10 Pa steps. Information about the operation mode was revealed by analyzing the spatial modulation of the time-averaged charged particle densities, i.e., we consider the plasma operating in the STR mode if the major ion densities inside the bulk are spatially modulated, and otherwise, it is considered to be operated in the non-STR mode. The results are summarized in figure 14 in the form of a "phase diagram" over the ($\phi, p$) parameter plane. The black crosses refer to the DA mode at 13.56 MHz and 18 MHz. The blue triangles correspond to the STR mode at both 13.56 MHz and 18 MHz while the red circles indicate the domain of a striated plasma (STR mode) at 13.56 MHz and a non-striated plasma (DA mode) at 18 MHz. The red and blue dashed curves mark the boundaries between the two modes at 13.56 MHz and 18 MHz, respectively. The boundaries can be approximated as hyperboles. For $f = 13.56$ MHz the $\phi p$ product (that defines the phase boundary) is found to be $\cong 1.1 \times 10^4$ V Pa (= const), while for $f = 18$ MHz, it acquires a value of $\cong 2.5 \times 10^4$ V Pa.

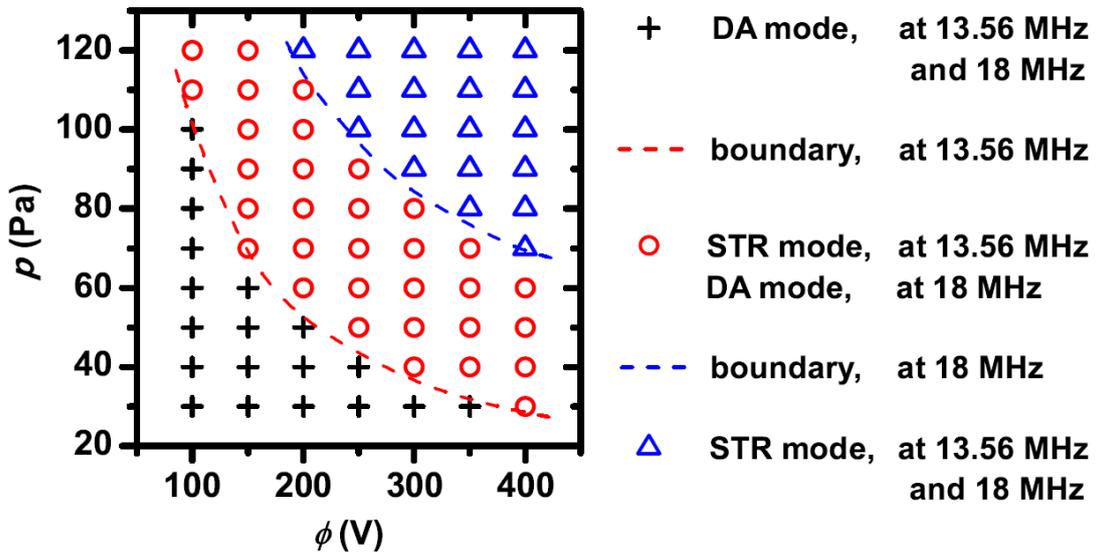

FIG.14. Phase diagram of the DA/STR discharge operation modes, as derived from PIC/MCC simulations, at $f = 13.56$ MHz and 18 MHz. The black crosses indicate conditions at which the discharge is operate in the DA-mode for both frequencies. The red circles indicate the parameter domain in which the plasma is striated (STR mode) at 13.56 MHz, but operated in the DA mode at 18 MHz. The blue triangles indicate plasma operation in the STR mode for both frequencies. The red and blue dashed lines represent the boundary between the two modes for 13.56 MHz and 18 MHz, respectively.

In order to discuss the mode transition between the DA and the STR modes induced by changing the gas pressure we analyze the spatio-temporal distributions of the ionization rate and the electric field, as well as the time-averaged profiles of the charged species densities shown in figure 15, at fixed values of the driving frequency (at 13.56MHz) and the voltage amplitude (at 250 V). At $p = 30$ Pa, the discharge is operated in a combination of $\alpha$ and DA modes. In this regime, the drift and ambipolar electric fields (see figure 15(b1)) accelerate electrons inside the bulk, leading to the ionization patterns shown in figure 15(a1) and to smooth ion density profiles (figure 15(c1)).

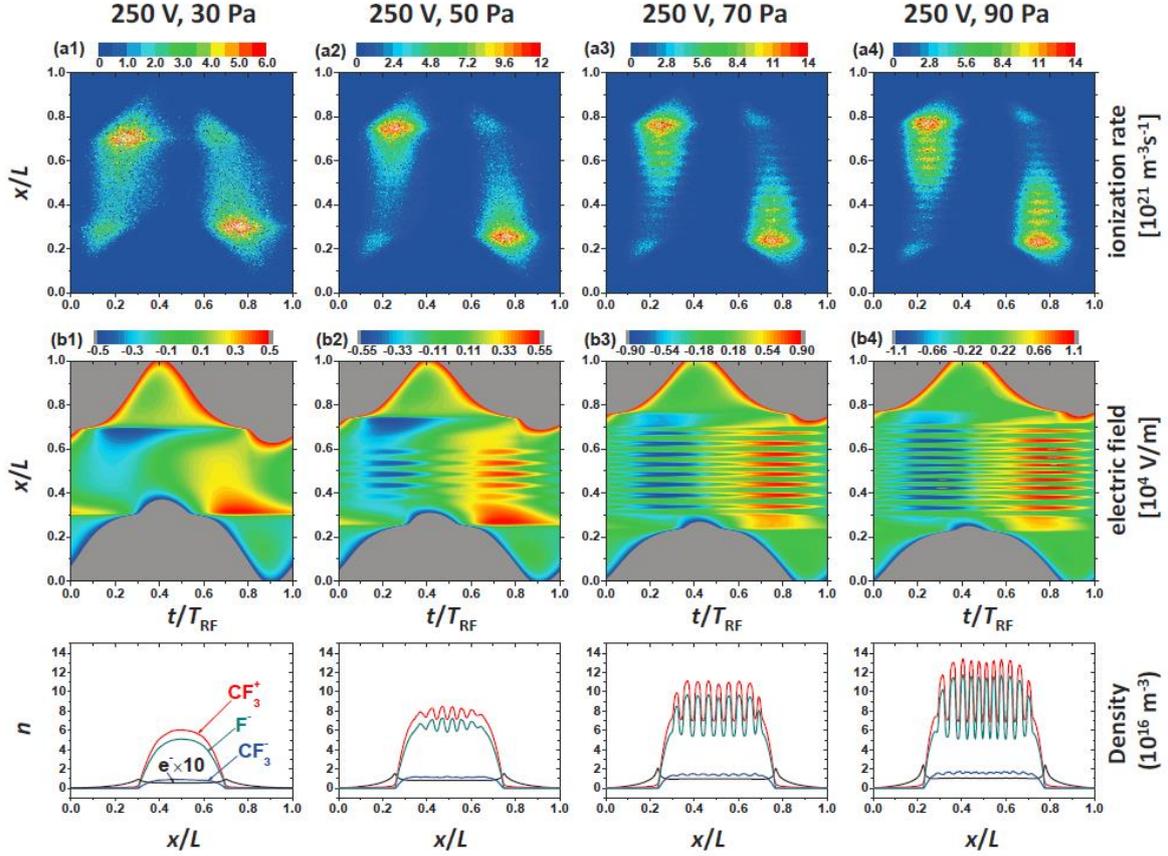

FIG. 15 PIC/MCC simulation results for the spatio-temporal distributions of the electron-impact ionization rate (first line), electric field (second line), and the axial profiles of the charged species time-averaged densities (third line) at pressure $p$ = 30, 50, 70, and 90 Pa. $f$ = 13.56 MHz, $L$ =1.5 cm, and $\phi$ = 250 V.

In this case the peak $CF_3^+$ and $F^-$ ion densities are lower than the minimum ion densities required to fulfill the resonance condition $\omega_-^2 + \omega_+^2 = \omega_{RF}^2$. Recall that due to their relatively low density $CF_3^-$ ions are not considered here. Under these conditions ($\phi$ = 250 V, 13.56 MHz) simulations at $p \geq$ 50 Pa yield a discharge operated in the STR mode. As the simulations are performed in 10 Pa steps we can only state that the pressure threshold for the DA to STR mode transition is between 40 Pa and 50 Pa. This threshold is required to reach high enough ion densities at a fixed voltage and frequency to fulfill the resonance condition. Figure 16 shows the maximum and minimum ion densities as a function of pressure for these discharge conditions. The minimum positive and negative ion densities at 50 Pa are marked as critical ion densities, $n_{+,c}$ and $n_{-,c}$, since 50 Pa is the lowest pressure at which the STR mode is observed in the simulation. At $p \geq 50$ Pa pressure the positive and negative ions respond to the alternating RF electric field and the plasma parameters (e.g., the electric field and the ionization rate) become modulated in space. As $p$ increases, the minima of $CF_3^+$ and $F^-$ ion densities slightly decrease and stabilize at two constants when $p > 100$ Pa, while the maxima of the ion densities increase (see figure 16). This is because in the STR mode the spatially modulated electric field can focus the positive and negative ions into the locations of the local ion density peaks, leading to the enhancement of the maximum ion density and the reduction of the minimum ion density. However, the minimum ion densities cannot become arbitrarily small because the ions then would not be able to respond to the alternating RF electric field.

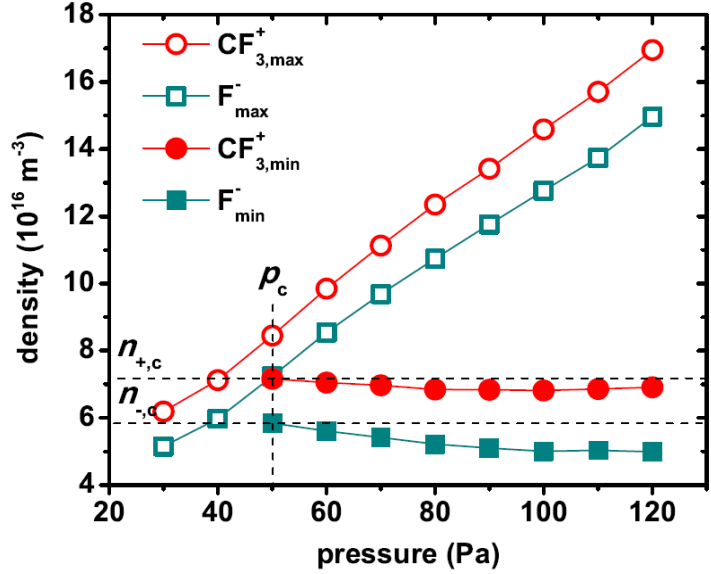

FIG. 16. The maximum and minimum of the $CF_3^+$ (circles) and $F^-$ (rectangles) ion densities as a function of $p$ for $\phi = 250$ V and $f = 13.56$ MHz. The other discharge conditions are the same as in figure 15. The horizontal dashed lines indicate the critical ion densities required for the formation of striations. The vertical dashed line indicates the pressure at which the mode transition occurs.

In figure 17, we illustrate the evolution of the spatio-temporal distribution of the ionization rate the electric field and the time-averaged axial profiles of the charged species densities with changing $\phi$, for $f = 13.56$ MHz and a fixed pressure of $p = 60$ Pa. The maximum and minimum of the $CF_3^+$ and $F^-$ ion densities as a function of the voltage amplitude are shown in figure 18. The mode transition induced by $\phi$ is quite similar to that induced by $p$. At $\phi = 100$ V and 150 V, the maximum ion densities are lower than the minimum ion densities required to fulfill the resonance condition and, thus, the discharge operates in the DA mode, while at 200 V the electric field and charged species densities already appear to be slightly modulated in space, since the ion densities are high enough. With a further increase of $\phi$, similarly to the case of constant voltage presented in figure 16, the minimum $CF_3^+$ and $F^-$ ion densities inside the bulk slightly decrease and tend to stabilize, while the maximum $CF_3^+$ and $F^-$ ion densities increase with $\phi$. In figure 18 the minimum positive and negative ion densities at 200 V are marked as critical ion densities, since for $\phi \geq 200$ V the STR mode is observed.

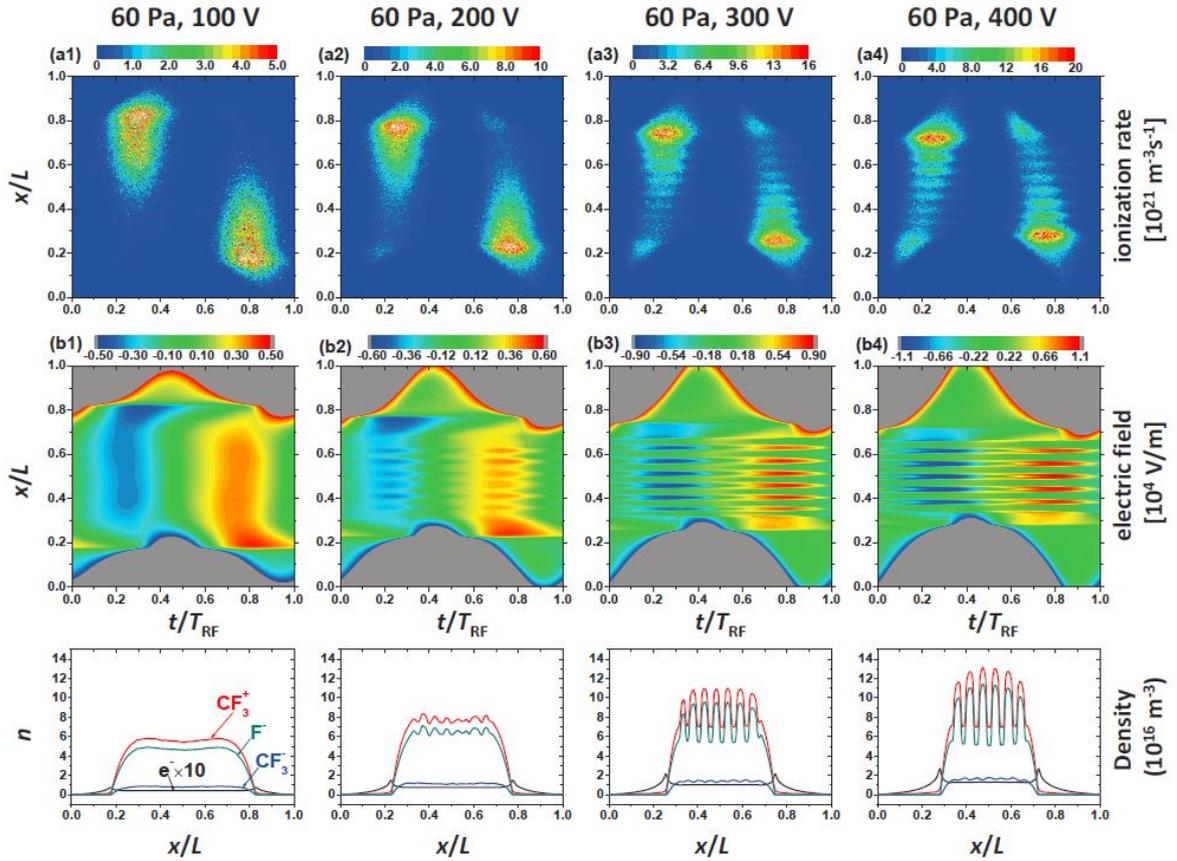

FIG. 17. PIC/ MCC simulation results for the spatio-temporal distributions of the electron-impact ionization rate (first line), electric field (second line), and the axial profiles of time-averaged charged species densities (third line) at voltage amplitudes $\phi$ = 100, 200, 300, and 400 V. $f$ = 13.56 MHz, $L$ = 1.5 cm, and $p$ = 60 Pa.

The phase diagram presented in figure 14 shows that the phase boundary line moves to higher pressure/voltages for higher driving frequencies, i.e., for 13.56 MHz the phase boundary is located at $\phi_P \cong 1.1 \times 10^4$ V Pa, while it is located at $\phi_P \cong 2.5 \times 10^4$ V Pa for 18 MHz. This shift is caused by the necessity of having higher ion densities to fulfill the resonance condition at higher driving frequencies. The densities can be increased both by increasing $\phi$ and/or $p$.

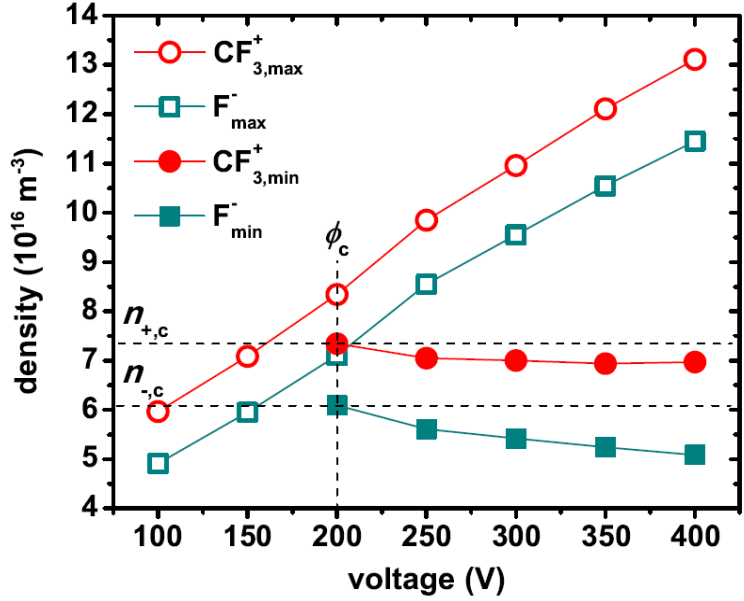

FIG. 18. The maximum and minimum of the $CF_3^+$ (circles) and $F^-$ (rectangles) ion densities as a function of $\phi$ for $p = 60$ Pa and $f = 13.56$ MHz. The other discharge conditions are the same as in figure 17. The horizontal dashed lines indicate the critical ion densities required for the formation of striations. The vertical dashed line indicates the voltage at which the mode transition occurs.

## IV. CONCLUSION

Self-organized striated structures of plasma parameters have been studied in capacitively coupled radio-frequency (CCRF) $CF_4$ plasmas by Phase Resolved Optical Emission Spectroscopy (PROES) and Particle-in-Cell/Monte Carlo Collision (PIC/MCC) simulations. These striations are found to be generated by the response of positive and negative ions to the alternating RF electric field in the bulk plasma, leading to a spatial modulation of the ion densities, the electric field, and, consequently, the ionization/excitation rate. Ions can respond to the RF electric field, if $\omega_+^2 + \omega_-^2 \geq \omega_{RF}^2$ (where $\omega_{+,-}$ is the positive/negative ion plasma frequency and $\omega_{RF}$ is the driving frequency). For various external parameters, the experimentally measured spatio-temporal distributions of the excitation rate are found to show good agreement with the ionization rate obtained from the PIC/MCC simulations.

The effects of changing the neutral gas pressure, the voltage amplitude, and the electrode gap on the striated structure of various plasma parameters have systematically been studied for different driving frequencies of 8 MHz, 13.56 MHz, and 18 MHz. At 8 MHz the discharge is found to be operated in this striation (STR) mode under all conditions studied, while transitions to the Drift-Ambipolar (DA) mode are observed as a function of these global control parameters at 13.56 MHz and 18 MHz.

In the STR mode, increasing the neutral gas pressure is found to lead to a smaller striation gap and a higher number of striations inside the plasma bulk. This is found to be caused by an increase of the transient bulk electric field as a function of the pressure due to a reduced electrical conductivity for higher collision frequencies. At higher electric fields electrons gain sufficient energy to sustain the striations by ionization over a shorter distance and, thus, the striation gap shrinks. Overall, the striation gap is determined by the necessity to balance charged particle losses via recombination by ionization locally in the vicinity of each striation, since ions are focused onto the adjacent striation by the transient and time averaged bulk electric field.

Increasing the driving voltage amplitude is found to lead to larger sheaths, a decrease of the bulk width, and an increase of the transient electric field in the bulk. While the striation gap remains constant as a function of the voltage, the number of striations decreases due to the decrease of the bulk width. The striated structure of the spatio-temporal distributions of the excitation/ionization gets more pronounced and more constricted to the vicinity of the collapsing sheath edge at higher voltages. While the first effect is caused by the increase of the transient electric field, the latter is caused by the decrease of the bulk width. Electrons cross multiple striations, while they are accelerated by the high bulk electric field from the expanding to the opposite collapsing sheath edge. A shorter bulk width reduces the length, over which they gain energy, and, thus, they reach high enough energies to cause excitation/ionization only close to the collapsing sheath edge at high voltages (short bulk widths).

The bulk width is found to be increased as a function of the electrode gap, while the striation gap remains approximately constant. Consequently, the number of striations inside the bulk increases as well. Electrons are accelerated from the expanding to the collapsing opposite sheath over a longer distance (bulk width) by high electric fields at larger electrode gaps. Thus, they reach the threshold energy for excitation/ionization at a larger distance from the collapsing sheath edge and excitation/ionization is observed over longer regions inside the bulk.

At 13.56 MHz and 18 MHz, transitions between the DA- and the STR-modes are found to be induced by changing the pressure and the RF voltage amplitude. A phase diagram over the ($\phi, p$) plane was constructed to demonstrate the parameter domains of DA and STR regimes. For a given frequency, the phase boundary is found to be approximated by a constant $p\phi$ product required to generate high enough ion densities to allow the ions to react to the RF electric field, i.e., to fulfill the resonance criterion. The critical $p\phi$ product is found to increase as a function of the driving frequency according to the resonance criterion, i.e. at higher driving frequencies higher ion densities are required to generate striations.

Generally, the striated discharge structure of electronegative CCRF plasmas is found to occur over wide parameter ranges of $f$, $p$, $L$, and $\phi$. Thus, the STR-mode is expected to have an important impact on the plasma-based applications, such as plasma enhanced chemical vapor deposition (PECVD), which is usually performed under similar conditions as analyzed in this work.

## ACKNOWLEDGMENTS

This work has been financially supported by the National Natural Science Foundation of China (NSFC) (Grants No. 11335004 and No. 11405018), by the Hungarian National Research, Development and Innovation Office, via Grant No. NKFIH 119357, by the German Research Foundation (DFG) within the frame of the collaborative research center SFBTR87, and by the US National Science Foundation (Grant No. 1601080).